\documentstyle[graphicx,times]{jaa}
\twocolumn
\begin{document}

\title[]{Variability of the Spectral Energy Distribution of the Blazar S5 0716+714}

\author[B. Rani, Alok C. Gupta and Paul J. Wiita]{B. Rani$^{1}$\thanks
{E-mail: bindu@aries.res.in}, Alok C. Gupta$^{1}$, Paul J. Wiita$^{2}$\\
$^{1}$Aryabhatta Research Institute of Observational Sciences (ARIES), Nainital, 263129, India \\
$^{2}$Department of Physics, The College of New Jersey, P.O.\ Box 7718, Ewing,
NJ 08628, USA }

\date{Accepted . Received ; in original form 2010 November 24}
\pagerange{\pageref{firstpage}--\pageref{lastpage}} \pubyear{2010}
\maketitle
\label{firstpage}

\begin{abstract}
The emission from blazars is known to be variable at all wavelengths. The flux
variability is often accompanied by spectral changes. Spectral energy 
distribution (SED) changes must be
associated with changes in the spectra of
emitting electrons and/or the physical parameters of the jet.
Meaningful modeling of blazar broadband spectra  is required to
understand the extreme conditions within the emission region.  Not only
is the broadband SED crucial, but also information about its variability is needed  
to understand how the highest states of emission occur and how they
differ from the low states. This may help in discriminating
between models. 
Here we present the results of our SED modeling of 
the blazar S5 0716+714 during various phases of its activity. The SEDs are classified 
into different bins depending on the optical brightness state of the source. 

\end{abstract}


\section{Introduction}
\label{sec:intro}
S5 0716+714 is a bright, high declination BL Lac object at a redshift, $z = 0.31 \pm$0.08 
 (Nilsson et al.\ 2008). This source has been extensively studied across the whole electromagnetic 
spectrum and exhibits strong variability on a wide range of timescales, ranging from minutes 
to years (e.g., Wagner et al.\ 1990; Heidt \& Wagner 1996; Villata et al.\ 2000; Raiteri et al.\ 2003; 
Montagni et al.\ 2006; Ostorero et al.\ 2006; 
Gupta et al.\ 2008a, b, 2009 and references therein). Nearly periodic oscillations  of $\sim$15 
minutes in the optical R band were detected in this source (Rani et al.\ 2010). The optical duty cycle 
of S5 0716+714 is nearly unity, indicating that the source is always in an active state in 
the visible (Wagner \& Witzel 1995).  This blazar was recently shown to be a strong source in 
the high energy gamma-ray band by Fermi-LAT (Abdo et al. 2009).

\section{Multi-frequency Data}
We carried out this study of the SEDs of the BL Lac object S5 0716$+$714 using high
quality multifrequency data in the literature. The data cover radio to optical energy 
bands. The optical U, B, V, R and I fluxes were collected from Raiteri et al.\ (2003), 
Montagni et al.\ (2006), Villata et al.\ (2000), Ostorero et al.\ (2006) and Gu et al.\  (2006). The data at 
radio frequencies of 22 and 37 GHz were taken from Salonen et al.\ (1987), Ter{\"a}sranta et al.\  (1992, 
1998, 2004, 2005). The UMRAO\footnote{http://www.astro.lsa.umich.edu/obs/radiotel/umrao.php} 
(University of Michigan Radio Astronomy Observatory) data at 4.8, 8 and 14.5 GHz frequencies 
was provided by Margo Aller. The data at optical frequencies spans over a period of 1994 
November through 2004 April while the radio data cover a time period of 1991 September to 
2005 May (Fig.\ 1).

We model the SEDs of S5 0716+714 using the above radio to optical frequency data. 
The long term data allow us to obtain six different mean SEDs of the source corresponding 
to different phases of its activity (Fig.\ 2). These six different SEDs  are characterized 
by different optical outburst phases and the radio data is  
averaged over the same time periods.

\begin{figure}
\begin{minipage}{80mm}
\centerline{\includegraphics[scale=.4]{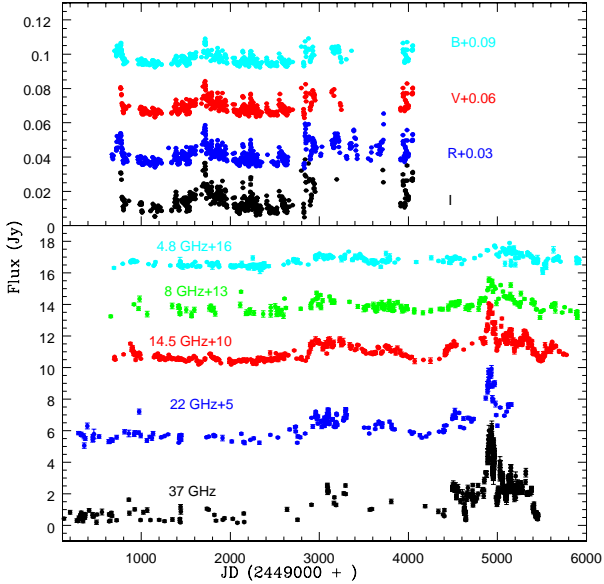}}
\caption{The multi-frequency radio and optical data of the BL Lac object 
S5 0716$+$714. }
\label{fig1}
\end{minipage}
\end{figure}

\section{Analysis and Results} 
\subsection{SED Modeling}
We use a homogeneous synchrotron 
self-compton model (SSC) with a broken power law (BPL) to fit the lower energy
part of observed spectra which are characterized by synchrotron spectra. The SED model fitting is 
achieved by using a  SED code available on-line\footnote{http://tools.asdc.asi.it/}. The best fit 
model was obtained by varying the parameters of a numerical SSC code (Tramacere 2007; Tramacere et al.\ 2009).
The model assumes that radiation is produced within a single zone of jet ($\sim$ radius R), moving
relativistically at a small angle to the line of sight of the observer. The observed radiation
is  amplified via the boosting factor $\delta$ = [$\Gamma (1 - \beta ~cos\theta)]^{-1}$.

\begin{figure}
\begin{minipage}{80mm}
\centerline{\includegraphics[scale=.4]{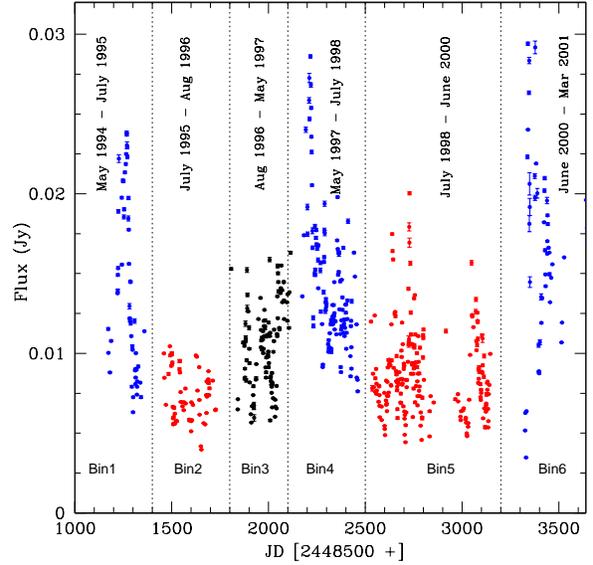}}
\caption{The optical R passband flux of  S5 0716$+$714 divided into six different bins 
on the basis of its optical activity. Bin1, Bin4 and Bin6 represent the 
optical outburst phases while Bin2, Bin3 and Bin5 represent dimmer phases of the 
source.                 }
\label{fig2}
\end{minipage}
\end{figure}

\begin{figure*}
\centerline{\includegraphics[scale=0.72]{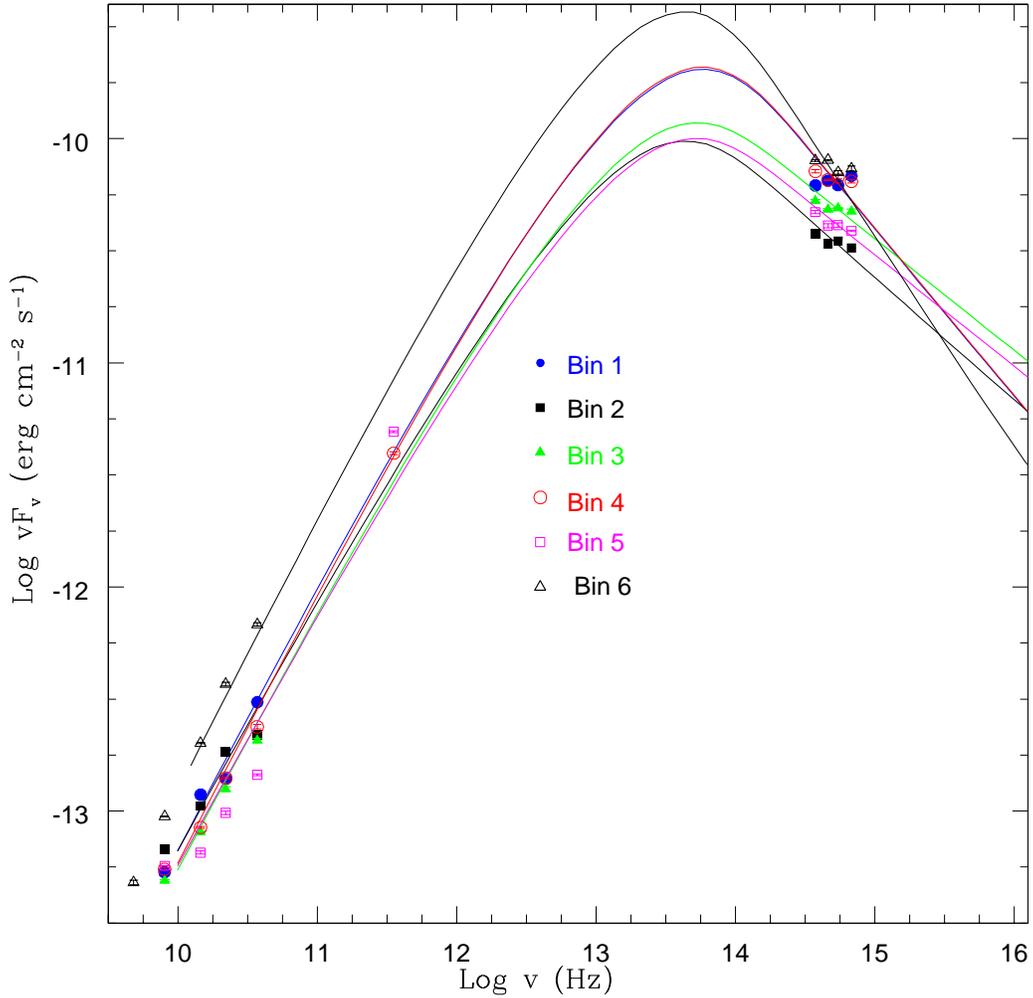}}
\caption{ The SSC modeled SEDs of the source corresponding to six different Bins.     }
\label{fig3}
\end{figure*}

\begin{table*}
\scriptsize
\caption{Fitted parameters of SED model}
\begin{tabular}{lccccccccccc} \\\hline
        &Log R  &   B       & $\delta$  & N           &  Log $\Gamma_{min}$  &   Log $\Gamma_{max}$  &$n_{1}$ & $n_{2}$ & Log $\Gamma_{b}$ & Log $\nu_{peak}$ & Log $\nu F_{\nu}$$_{peak}$ \\
     &  (cm)  &   (Gauss)  &            &  (cm$^{-3}$) &      &        &        &          &        & (Hz)   & (erg cm$^{-2}$ s$^{-1}$)         \\\hline
Bin1 &  16.8  &       0.4  &      12.8   &      25       &     0.2     &          6  &     0.4   &    4.5   &     3.20   &13.76 &-9.69    \\
Bin2 &  16.8  &       0.4  &      11.5   &      30       &     0.2     &          6  &     0.6   &   4.1    &     3.14   &13.61 &-10.01     \\
Bin3 &  16.8  &       0.4  &      11.0   &      30       &     0.2     &          6  &     0.5   &   4.0    &     3.18   &13.75 &-9.92     \\
Bin4 &  16.8  &       0.4  &      12.5   &      25       &     0.2     &          6  &     0.3   &   4.5    &     3.20   &13.76 &-9.67    \\
Bin5 &  16.8  &       0.4  &      11.0   &      30       &     0.2     &          6  &     0.6   &   4.0    &     3.18   &13.72 &-9.99     \\
Bin6 &  16.8  &       0.4  &      16.0   &      25       &     0.2     &          6  &     0.2   &   4.9    &     3.20   &13.64 &-9.42    \\\hline
\end{tabular}\\
R : Size of emitting region  \\
B     : Magnetic field       \\
$\delta$  : Doppler boosting factor   \\
N         : Number density of emitting electrons \\
$\Gamma$$_{min}$, $\Gamma$$_{max}$ : Minimum and maximum values of Lorentz factor \\
$\Gamma$$_{b}$ : Lorentz factor corresponding to break energy of electrons \\
$n$$_{1}$, $n$$_{2}$ : Power law indices  \\
$\nu_{Peak}$ : Synchrotron peak frequency \\

\end{table*}

Below and above the peaks, the spectrum can
be approximated with power law profiles with indices $\alpha_{1}$ and $\alpha_{2}$, respectively. The power law
spectra in AGNs are naturally produced if the emitting electron follow a power-law distribution
in energy. We approximate this behaviour with a broken power law (BPL) with indices $n_{1}$ and
$n_{2}$, respectively below and above the break energy $\Gamma_{b}$m$_{e}$c$^{2}$ :

\begin{equation}
P(\gamma )=\left\{ \begin{array}{ll}
                    N\Gamma ^{-n_1}  &  \mbox{if $\Gamma < \Gamma _b$} \\
                    N\Gamma _b^{n_2-n_1} \Gamma ^{-n_2}  &  \mbox{if $\Gamma > \Gamma _b$}
                   \end{array}
           \right.              
\end{equation}

With these approximations, we can completely specify the model with the following parameters:
magnetic field intensity (B), size of emission region (R), Doppler boosting factor
($\delta$), power-law indices (n$_{1}$,  n$_{2}$),  number density of emitting
electrons (N) and Lorentz factor of the electrons
at the break energy ($\Gamma_{b}$). The values of all these fitted parameters for
different SEDs are listed in Table 1.

\section{Discussion and Conclusions}
\subsection{Limitations to the model}
Although, as seen in Fig.\ 3, we were able to achieve reasonably good fits of the synchrotron
emission for all six SEDs of the source, one should bear in mind that the one-zone BPL
model is  over-simplified in  accounting for the radio-optical blazar emission. The
applicability of a single-zone emitting region has been questioned by a number of authors
(e.g., Vittorini et al.\ 2009, Raiteri et al.\ 2010). They showed that the BL Lac SED can 
be more successfully modeled with two synchrotron components (two different emitting populations).

Furthermore, unfortunately, we do not have synchrotron peak measurements for the
the source, which would significantly help to constrain the model. 
Another limitation to the model is that we do not correct our optical measurements for any
possible host galaxy contribution. Some objects, such as BL Lac itself, have a significant
contribution from starlight to the optical bands, which will modify the calculated synchrotron emission in this
region, especially during the low states. We also do not make any corrections for Galaxy or
internal absorption, which may again slightly affect the optical-UV part of the spectra.
Last, but not least, we stress that our data is not strictly simultaneous but has been averaged over
a period of months. As blazars are highly variable over 
timescales of a day or less, the time differences and averaging might have compromised somewhat the modeling.

\subsection{Spectral Energy Distribution variation}
To attempt a study of how the physical parameters related to emission region 
and synchrotron emission are changed when the BL Lac S5 0716+714 goes 
through various phases of its activity we used the simplest approach by 
fitting  a single-zone SSC model. The observed results can be summarized as : \\
1. No change between the R and B bands in the modeled SED are seen
during different phases of activity. \\
2. The Doppler boosting factor $\delta$ is higher during the optically 
bright states of the source compared to the dimmer phases of activity. \\
3. The number density (N) of electrons emitting synchrotron photons is 
larger when the source is in lower states. \\
4. The synchrotron peak frequency ($\nu_{peak}$) and peak intensity 
($\nu F_{\nu}$$_{peak}$) are comparatively higher during the 
optical outburst phases of the BL Lac object.  \\

This research was supported by CSIR Foreign travel grant {\bf Ref No.  TG/5295/1-HRD} and
 has made use of data from the University of Michigan Radio Astronomy Observatory
which has been supported by the University of Michigan and by a series of grants from the National
Science Foundation, most recently AST-0607523. BR is very thankful to Margo Aller for providing the
radio frequency data.

\bibliography{}

\begin{thebibliography}{}

\bibitem{abdo2009}
{Abdo}, A.~A., et al. 2009, ApJ, 707, 1310

\bibitem{gupta2008a}
{Gupta}, A.~C., et~al. 2008a, AJ, 136, 2359

\bibitem{gupta2008b}
{Gupta}, A.~C., {Fan}, J.~H., {Bai}, J.~M.,  \& {Wagner}, S.~J. 2008b, AJ,
  135, 1384

\bibitem{gupta2009}
{Gupta}, A.~C., {Srivastava}, A.~K.,  \& {Wiita}, P.~J. 2009, ApJ, 690, 216


\bibitem{2006A&A...450...39G} Gu, M.~F., Lee, C.-U., Pak, S., Yim, H.~S., \& 
Fletcher, A.~B.\ 2006, A\&A, 450, 39 


\bibitem{heidt1996}
{Heidt}, J.,  \& {Wagner}, S.~J. 1996, A\&A, 305, 42


\bibitem{2006A&A...451..435M} Montagni, F., Maselli, A., Massaro, E., Nesci, R., 
Sclavi, S., \& Maesano, M.\ 2006, A\&A, 451, 435 


\bibitem{2008A&A...487L..29N} Nilsson, K., Pursimo, T., Sillanp{\"a}{\"a}, 
A., Takalo, L.~O., \& Lindfors, E.\ 2008, A\&A, 487, L29 


\bibitem{2006A&A...451..797O} Ostorero, L., et al.\ 2006, A\&A, 451, 797 



\bibitem{raiteri2003}
{Raiteri}, C.~M., et~al. 2003, A\&A, 402, 151


\bibitem{2010arXiv1009.2604R} Raiteri, C.~M., et al.\ 
2010, arXiv:1009.2604 


\bibitem{2010ApJ...719L.153R} Rani, B., Gupta, A.~C., 
Joshi, U.~C., Ganesh, S., \& Wiita, P.~J.\ 2010, ApJ, 719, L153 



\bibitem{1987A&AS...70..409S} Salonen, E., et al.\ 1987, A\&AS, 70, 409 


\bibitem{2009A&A...501..879T} 
Tramacere, A., Giommi, P., Perri, M., Verrecchia, F., \& Tosti, G.\ 2009, A\&A, 501, 879 


\bibitem{2007A&A...466..521T} 
Tramacere, A., Massaro, F., \& Cavaliere, A.\ 2007, A\&A, 466, 521 


\bibitem{1992A&AS...94..121T} Ter{\"a}sranta, H., et al.\ 1992, A\&AS, 94, 121 

\bibitem{1998A&AS..132..305T} Ter{\"a}esranta, H., et al.\ 1998, A\&AS, 132, 305 


\bibitem{2004A&A...427..769T} Ter{\"a}sranta, H., et al.\ 2004, A\&A, 427, 769 


\bibitem{2005A&A...440..409T} Ter{\"a}sranta, H., Wiren, S., Koivisto, P., Saarinen, 
V., \& Hovatta, T.\ 2005, A\&A, 440, 409 




\bibitem{villata2000}
{Villata}, M., et~al. 2000, A\&A, 363, 108


\bibitem{2009ApJ...706.1433V} Vittorini, V., et 
al.\ 2009, ApJ, 706, 1433 




\bibitem{wagner1990}
{Wagner}, S., {Sanchez-Pons}, F., {Quirrenbach}, A.,  \& {Witzel}, A. 1990,
  A\&A, 235, L1


\bibitem{wagner1995}
{Wagner}, S.~J.,  \& {Witzel}, A. 1995, ARA\&A, 33, 163



\end{thebibliography}

\label{lastpage}

\end{document}